\documentstyle[12pt]{article}

\begin{document}

\hfill \vbox{\hbox{UCLA/95/TEP/25}
             \hbox{hep-lat/9508010} }

\begin{center}

{\Large \bf SO(3) vortices and disorder in the 2d SU(2) chiral
model}\footnote{Research supported in part by NSF Grant PHY89-15286} \\[2cm]

{\bf Tam\'as G.\ Kov\'acs}\footnote{e-mail address: kovacs@physics.ucla.edu}
 and {\bf E.\ T.\ Tomboulis}\footnote{e-mail address:tomboulis@uclaph.bitnet}
\\
{\em Department of Physics \\
University of California, Los Angeles \\
Los Angeles, California 90024-1547} \\[2cm]

\end{center}

\section*{Abstract}

We study the correlation function of the 2d SU(2) principal chiral model
on the lattice. By rewriting the model in terms of Z(2) degrees
of freedom coupled to SO(3) vortices we show that the vortices play a crucial
role in disordering the correlations at low temperature. Using a series
of exact transformations we prove that, if satisfied, certain inequalities
between vortex
correlations imply exponential fall-off of the correlation function at
arbitrarily low temperatures. We also present some Monte Carlo evidence
that these correlation inequalities are indeed satisfied.
Our method can be easily translated to the language of 4d SU(2) gauge
theory to establish the role of corresponding SO(3) monopoles in maintaining
confinement
at small couplings.

\vfill

\pagebreak

\paragraph { } It is commonly believed that two-dimensional
lattice spin models with a continuous nonabelian symmetry
exhibit no phase transitions at finite temperature. Indeed,
according to the Mermin-Wagner theorem, these models do not
have an ordered low temperature phase with spontaneous
breakdown of the continuous symmetry. However, this does not
in itself rule out the possibility of a Kosterlitz-Thouless (KT)
type phase transition. The KT transition is characterised by
power-law decay of the correlation function in the low
temperature phase as opposed to an exponential decay
above the critical temperature. The absence of this KT-type phase transition
has not been proved rigorously for nonabelian spin models.

Two dimensional spin models are well known to have properties analogous to
that of four dimensional gauge theories. In 4d nonabelian gauge theories
the corresponding problem is whether the system remains confining down to
arbitrarily small finite couplings. A 4d SU(2) lattice gauge theory can be
rewritten in terms of
SO(3) and Z(2) variables. This exhibits SO(3) monopoles in the measure, and
recently a program has been developed for establishing the presence of these
monopoles, and their associated strings, as a sufficient mechanism for
confinement at arbitrarily
small positive coupling \cite{Tomboulis}. Completely analogous
considerations can be developed for the SU(2)$\times$SU(2) chiral spin
model in 2d. In a previous paper \cite{Kovacs} we have rewritten the partition
function
and the two-point correlation function of this model in terms of SO(3) and
Z(2) variables, and presented some arguments supporting the role that
SO(3) vortices \cite{Kogut} (the analogues of gauge monopoles) and strings
connecting them can play in disordering the system.

In the present paper we use this picture to derive an exponentially falling
upper bound on the correlation function of the SU(2)$\times$SU(2) model
in terms of correlations between SO(3) ``stringy'' vortices.
The main idea is the
following. After rewriting the SU(2) degrees of freedom in terms of Z(2) and
SO(3) variables and performing a duality transformation on the Z(2) variables
the resulting system is an Ising model on the dual lattice coupled to SO(3)
vortices. The effect of the vortices on the Ising spins is similar to that of
an external magnetic field, provided that appropriate expectations of vortices
pairwise connected by long strings winding around the lattice satisfy
certain positivity conditions and remain nonvanishing in the large lattice
limit. It is crucial that these expectations are defined with respect to an
SO(3) (rather than an SU(2)) measure (see below). One thus may bound the
correlation function of the original SU(2) model from above by a corresponding
expectation  in an effective
Ising model with a nonzero external magnetic field.
This correlation in turn is well known to decay exponentially
thus giving an exponentially falling upper bound to the  two-point
function of the SU(2) principal chiral model.

We shall work on a finite two dimensional square lattice $\Lambda$ with
periodic boundary conditions. The degrees of freedom (``spins'') are SU(2)
group elements $U_s$ attached to lattice sites with nearest neighbour
ferromagnetic couplings. The partition function of the system is
\begin{equation}
 Z = \prod_{s \in \Lambda} \int dU_s
 \hspace{2mm} \exp \left[ \beta \sum_{l \in \Lambda}
 \mbox{tr} U_l \right],
     \label{eq:PF}
\end{equation}
where $\beta$ is the inverse temperature, sites, links and plaquettes are
labelled with $s$, $l$, $p$ respectively, and $U_l=U^\dagger_s U_{s'}$ with
$[ss']$ being
$\partial l$, the ordered boundary of the link $l$. Throughout the paper all
group integrations (discrete and continuous) will be performed using the
Haar-measure normalised to unity.

Let $\{ \tau_1, \tau_2 \} $ be a pair of Z(2) elements, Z(2)=$\{ \pm 1 \} $
being
the centre of the symmetry group SU(2). We can introduce a twist $\tau_1$
in the `1' direction and $\tau_2$ in the other and denote by $Z(\tau_1,\tau_2)$
the partition sum in the presence of these twists. By a twist in the direction
$i$ we mean that on a stack of links winding around the lattice along
the $i$ direction the couplings are changed from $\beta$ to $\tau_i \beta$.
Physically $\tau_i=-1$ means that a topologically nontrivial ``domain-wall''
was created along the affected links. This domain-wall winds around the
lattice,
it is closed but not a boundary of any region. The order parameter that we
shall consider is the expectation of such a twist defined as
\begin{equation}
 G(L) = \frac{\int d \tau_1 \, \int d \tau_2 \, \tau_2 Z(\tau_1,\tau_2)}
             {\int d \tau_1 \, \int d \tau_2 \,        Z(\tau_1,\tau_2)}=
 \frac{Z_+(L) - Z_-(L)}{Z_+(L) + Z_-(L)},
     \label{eq:GL}
\end{equation}
where $Z_{\pm}(L)=\int d \tau_2 \; Z(\pm 1,\tau_2)$ and $L$ is the linear
size of the lattice. (\ref{eq:GL}) is the spin analog of the electric-flux
free-energy order parameter of gauge theory \cite{'tHooft}. If by sending the
lattice size to infinity, $G(L)$
goes to zero exponentially, the system is essentially
disordered and it does not ``feel'' the presence of the enforced
domain-wall even when the lattice becomes infinitely large.

$G(L)$ has qualitatively the same asymptotic behaviour as the spin-spin
correlation function, and it can indeed be rigorously proved \cite {TY}
that the two-point correlation function for separation
$L$ is bounded from above by a constant times $G(L)$. This means that
the exponential fall-off of $G(L)$ would imply the same asymptotic behaviour
of the spin-spin correlation function.

For a quantitative study of $G(L)$ we rewrite the theory (\ref{eq:PF}) by means
of a decomposition into Z(2) and SU(2)/Z2 $\approx$ SO(3) variables
as was done in \cite{Kovacs}, to which we refer for details.
After performing an additional duality transformation on the Z(2) degrees of
freedom the order parameter assumes the form
\begin{eqnarray}
  G(L) = \frac{1}{Z_+(L)+Z_-(L)} \;
  \prod_{s \in \Lambda} \int dU_s \,
 \exp \left( \sum_{l \in \Lambda} M(U_l) \right) \hspace{5cm}
     \nonumber \\
 \times \prod_{p \in \Lambda} \int d\omega_p \;
 \chi_{d \eta_p}(\omega_p) \; \eta_C \;
 \exp \left[ \sum_{l \not{\in} C} K(U_l) \delta \omega_l
 - \sum_{l \in C} K(U_l) \delta \omega_l  \right],
     \label{eq:2pt}
\end{eqnarray}
where $\eta_l= \mbox{sign} \, \mbox{tr} U_l$, d$\eta_p=\prod_{l \in \partial p}
\eta_l$, the $\chi$'s are characters of Z(2) and $\eta_C=\prod_{l \in C}
\eta_l$
with $C$ being a loop winding around the lattice in the direction perpendicular
to the twist. The functions $M(U_l)$ and $K(U_l)$ are given by
\begin{equation}
 K(U_l) = \frac{1}{2} \ln \coth \beta | \mbox{tr} U_l|, \hspace{6mm}
 M(U_l) = \frac{1}{2} \ln \left( \cosh \beta | \mbox{tr}U_l |
 \sinh \beta | \mbox{tr} U_l | \right) \label{eq:dual}
\end{equation}
Now the integrand in (\ref{eq:2pt}) depends on the SU(2) variables $U_s$ only
through the SU(2)/Z(2) cosets since it is clearly invariant under a local
transformation of the $U_s$'s by elements of Z(2). Thus the integration is
effectively over SO(3) degrees of freedom. In compensation, (\ref{eq:2pt})
also contains Z(2) spins $\omega_p$
attached to plaquettes. Spins on neighbouring plaquettes sharing the link $l$
interact via the
fluctuating coupling $K(U_l)$. The Z(2) part of the system is now essentially
a ferromagnetic Ising model but with couplings
depending on the SO(3) degrees of freedom. Because of the duality
transformation,
low temperature in the original model corresponds to high temperature
i.e.\ small $K(U_l)$'s in this Ising model. The Ising and the SO(3)
variables are further coupled through the Z(2) characters.
By definition $\chi_{d \eta_p} (\omega_p)=1$ \ if d$\eta_p=1$: and $\chi_{d
\eta_p} (\omega_p) = \omega_p$ \ if d$\eta_p=-1$, which
means that there is an SO(3) vortex on the plaquette $p$. The Z(2)
characters thus couple the Ising spins to SO(3) vortices. It is also
obvious from the construction that vortices only occur pairwise connected by
"$\eta$-strings", i.e. stacks of links on which $\eta_l=-1$ (Fig.\ 1).
This terminology is motivated by the analogy with
Dirac strings in gauge theories. The strings are not gauge invariant but their
endpoints, the vortices (monopoles in gauge theories), are.
Finally, the $M(U_l)$ part of the action depends only on the SO(3)
variables.

In this representation $G(L)$ involves a twist along $C$ in the
{\em dual} Ising model of $\omega$ spins. Notice that while in the original
system the twist was
along a stack of links winding around the lattice in the ``1'' direction,
in the dual system the twisted links form a loop $C$ going around the
lattice in the direction perpendicular to ``1''.
Furthermore, there are two additional crucial factors in the measure:
$\eta_C$, and the product of Z(2) characters that couple SO(3) vortices
to Z(2) spins. For a fixed SO(3) configuration with vortices at
$(p_1,...p_{2n})$ and $S$ $\eta$-strings crossing $C$, these give an
overall factor of $(-1)^S \prod_{i=1}^{2n} \omega_{p_i}$. Were it not for
this additional factor depending on the SO(3) configuration, the system
would essentially be an ordinary ferromagnetic Ising model at high temperature
with unbroken
Z(2) symmetry and $G(L)$ going to unity on large lattices.

Let us then look at the Z(2) characters in the measure. In a fixed
``background''
SO(3) configuration containing $2n$ vortices, the sum over the Ising
configurations gives a $2n$-point function of the Ising system with couplings
$K(U_l)$. When finally the SO(3) variables are integrated out, we get a sum
over all possible (even) numbers and locations of vortices which translates
into a sum of all possible correlations in the Ising system taken with
additional weights coming from the SO(3) part of the measure. This is very
much reminescent of the expansion of the partition function of an Ising
model with respect to an external magnetic field, where the same type
of sum appears. This motivates the following physical picture. The high
temperature Ising system of $\omega$ spins would be in its symmetric phase
by itself but the coupling to the vortices breaks the symmetry by creating
an effective external magnetic field. In this broken phase the free energy
of the twist (\ref{eq:2pt}) along $C$ grows exponentially with the lattice size
implying exponential decay for $G(L)$.

To make this argument quantitative we proceed as follows. We first
insert a delta function in the measure in (\ref{eq:2pt}) to constrain
$\eta_C$ to unity. Let $G(L,C_+)$ denote $G(L)$ computed in the presence of
this constraint. It can be rigorously proven that $G(L) \leq G(L,C_+)$,
so it is enough to verify exponential fall-off for $G(L,C_+)$. We next
compare $G(L,C_+)$ to the quantity
\begin{equation}
G_{\mbox{\tiny eff}}(h,L) = \frac{1}{Z_{\mbox{\tiny eff}}(h,L)}\prod_{p \in
\Lambda} \int d\omega_p \; \exp \left[ \sum_{l \not{\in} C} K({\bf 1}) \delta
\omega_l  - \sum_{l \in C} K({\bf 1} ) \delta \omega_l + \sum_{p\in \Lambda}
h\omega_p \right], \label{eq:eff}
\end{equation}
i.e. the expectation of a twist along $C$ in an effective Ising system of
$\omega$ spins coupled to a magnetic field $h$. Now using a method similar to
Ginibre's proof of Griffiths-type inequalities \cite{Ginibre}, the difference
$G_{\mbox{\tiny eff}}(h,L)-G(L,C_+)$ can be expressed as a sum of terms,
each term being of the form
\begin{equation}
 \Gamma_{L,\bar{\theta}}[P_1,P_2,...,P_n] \equiv
 \langle \prod_{i=1}^{n} (\theta^-_{P_i} - \bar{\theta} \theta^+_{P_i})
 \prod_{p \not{=} \{P_i\}} \theta^+_p \rangle_L,
     \label{eq:vor}
\end{equation}
times positive numerical coefficients. Here, $P_1, ..., P_n$ is a set of
n vortex pairs,
$\theta^{\pm}_{P_i}=\theta^{\pm}_{p_{i1}}\theta^{\pm}_{p_{i2}}$,
where $p_{i1},p_{i2}$ denote the locations of the two vortices of the $i$-th
pair, and $\theta^{\pm}_p$ constrains d$\eta_p$ to be $\pm 1$. $\langle \;-\;
\rangle_L$ means integration with respect {\it only to a pure SO(3) measure}
defined
simply by the action
$M(U_l)$, \ref{eq:dual}, and
including the constraint $\eta_C=1$. Also, we defined $\bar{\theta}\equiv \tanh
h$.
Since the integration measure is positive, at $\bar{\theta}=0$ expression
(\ref{eq:vor}) is positive, hence  $G_{\mbox{\tiny eff}}(0,L)\leq G(L)$. By
continuity the same then holds in a neighbourhood of $\bar{\theta}=0$ ($h=0$).
If this neighbourhood does not shrink to zero when the lattice
size is sent to infinity, $G(L)$ must obey exponential decay since
$G_{\mbox{\tiny eff}}(h,L)$ does for any non-zero $h$, the mass-gap being
proportional to $\tanh h$. We have thus rigorously reduced the existence of
a mass gap to a condition on vortex pair correlations.

Next consider the assertion
\begin{equation}
 \Gamma_{L,\bar{\theta}}(P_1,P_2,...,P_{n}) \; \geq \;
 \mbox{const.}  \times  \Gamma_{L,\bar{\theta}}(P_1) \times ...
  \times \Gamma_{L,\bar{\theta}}(P_n),
     \label{eq:factorize}
\end{equation}
i.e. that the correlators $\Gamma$ of vortex pairs (\ref{eq:vor}) are bounded
by products of vortex pair correlations. This (highly nontrivial) inequality
can
be rigorously proven for $\bar{\theta}=0$ by an argument that reduces
(\ref{eq:factorize}) to an application of the FKG inequalities \cite{FKG}. It
is then
very plausible that it also holds for sufficiently small values of
$\bar{\theta}$.
Indeed, preliminary results indicate that (\ref{eq:factorize}) holds for
$\bar{\theta}$ such that the r.h.s. remains nonnegative, i.e. for
$\bar{\theta} \leq  k\bar{\theta}_0$, where $k$ is a numerical constant of
order unity, and $0\leq \bar{\theta}_0$ such that
\begin{equation}
 \Gamma_{L,\bar{\theta}}(P) = \langle \theta^-_{p_1}
 \theta^-_{p_2}  \prod_{p \not{=} p_i} \theta^+_p \rangle_L \;
 -  \; \bar{\theta}_0 \langle \prod_{p \in \Lambda} \theta^+_p \rangle_L
 \geq 0
\end{equation}
holds on arbitrarily large lattices regardless the
location ($p_1$ and $p_2$) of the two members of the vortex pair $P$. This is
equivalent to the
statement that the free energy cost of a pair of vortices
\begin{equation}
 F_L(p_1,p_2) = - \frac{1}{\beta} \ln
 \frac{ \langle \theta^-_{p_1} \theta^-_{p_2}
 \prod_{p \not{=} p_i} \theta^+_p \rangle_L}{ \langle
 \prod_{p \in \Lambda} \theta^+_p \rangle_L}
     \label{eq:vor_energy}
\end{equation}
is bounded for any lattice size. The only case when this free energy can
in principle be sensitive to the lattice size is when $p_1$ and $p_2$ are
on opposite sides of $C$ and the constraint $\eta_C=1$ forces the
$\eta$-string connecting them to go around the lattice (Fig.\ 1).

Now {\it in two dimensions} the energy of such an $\eta$-string stays
finite as $L \rightarrow \infty$ in the semiclassical approximation, where one
obtains $F_L(P)= {\mbox{constant}}$, whereas it diverges with $L$ in higher
dimensions. This is due to flux spreading \cite{Yaffe} that allows the cost of
the string creation to be spread laterally in the direction perpendicular to
the string. As explained above, constant $F_L(P)$ implies $h$ and hence a
mass gap proportional to $\exp{-({\mbox{const}})\beta}$. In the remainder
of the paper we shall briefly present the results of a Monte Carlo
measurement of $F_L(p_1,p_2)$ that indicate that this behavior of $F_L(P)$
holds for the exact expectation (\ref{eq:vor_energy}) . Details
of this Monte Carlo calculation will appear elsewhere.

We measured the quantity $\exp(-\beta F_L(p_1,p_2))$ by Monte Carlo using
the SO(3) action $|\mbox{tr}U_l|$ appearing in (\ref{eq:vor_energy}).
The simulations were performed on lattices $5 \le L \le 13$ at $\beta=2.0$.
As can be seen from the location of the specific heat peak, this is
already on the weak coupling side of the crossover in the SO(3) model.
The measurement was done by simply counting in a long Monte Carlo run the
number of configurations having exactly two vortices at the fixed locations
$p_1$ and $p_2$ and in addition satisfying the constraint $\eta_C=1$
(Fig.\ 1). This constraint forces the eta string connecting the two
nearby vortices to run all the way around the lattice. Finally the number
of these configurations was divided by the number of configurations
containing no vortices at all. Our results are summarised in Figure 2.
We can see that the probabilty of having a vortex pair at a given location
with a long eta string decreases on small lattices until it stabilises on
moderate sized lattices $(L \approx 8-9)$ at a nonzero constant value.
Recall that for our purposes it is enough that this quantity remains nonzero in
the
$L \rightarrow \infty$ limit. Figure 3 illustrates the pronounced effect on
(\ref{eq:vor_energy}) of flux spreading in the lateral direction.

To summarise, we have seen how vortices can disorder
the correlation function in the two dimensional SU(2)$\times$SU(2) chiral
spin model at arbitrarily low temperatures. By an exact rewriting
of the original model to separate SO(3) and Z(2) degrees of freedom, we derived
sufficient conditions on certain SO(3) vortex correlations for the existence of
an exponentiallyfalling upper bound to the order paramete. To complete the
argument we made two interrelated assumptions concerning the behavior of the
correlations (\ref{eq:factorize}), (\ref{eq:vor_energy}).
We presented the results of
a Monte Carlo calculation that confirm the expected behavior of
(\ref{eq:vor_energy}).
Clearly it would be very worthwile to find an analytic proof of these two
assumptions thus completing a rigorous demonstration of the absence of a
KT-type phase transition in this model. It would be also interesting to extend
these arguments to similar models with different symmetry groups. There are
two obvious possible ways of generalisation. One is to SU(n) principal chiral
models and the other to O(n) vector models by noting that the SU(2)
principal chiral model is equivalent to the O(4) vector model.

\section*{Acknowledgements}

We thank the Department of Theoretical Physics, Kossuth Lajos University,
Debrecen, Hungary for granting us part of the computer time used for
the Monte Carlo simulation.

\pagebreak

\section*{Figure captions}

\begin{tabular}{r p{11cm}}

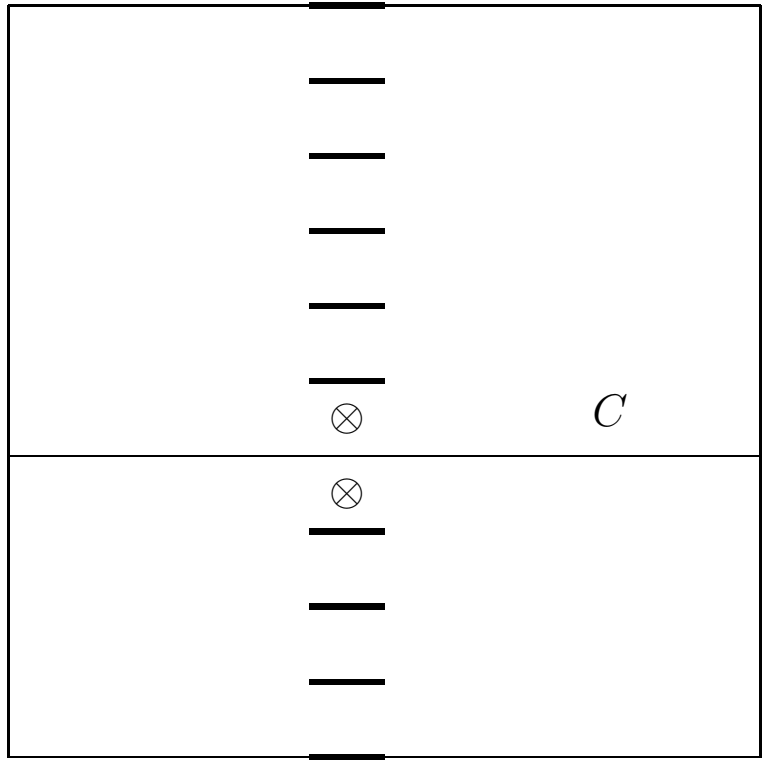
{\em Figure 1} &  Two vortices connected by an $\eta$-string winding
around the lattice.  \\[2mm]

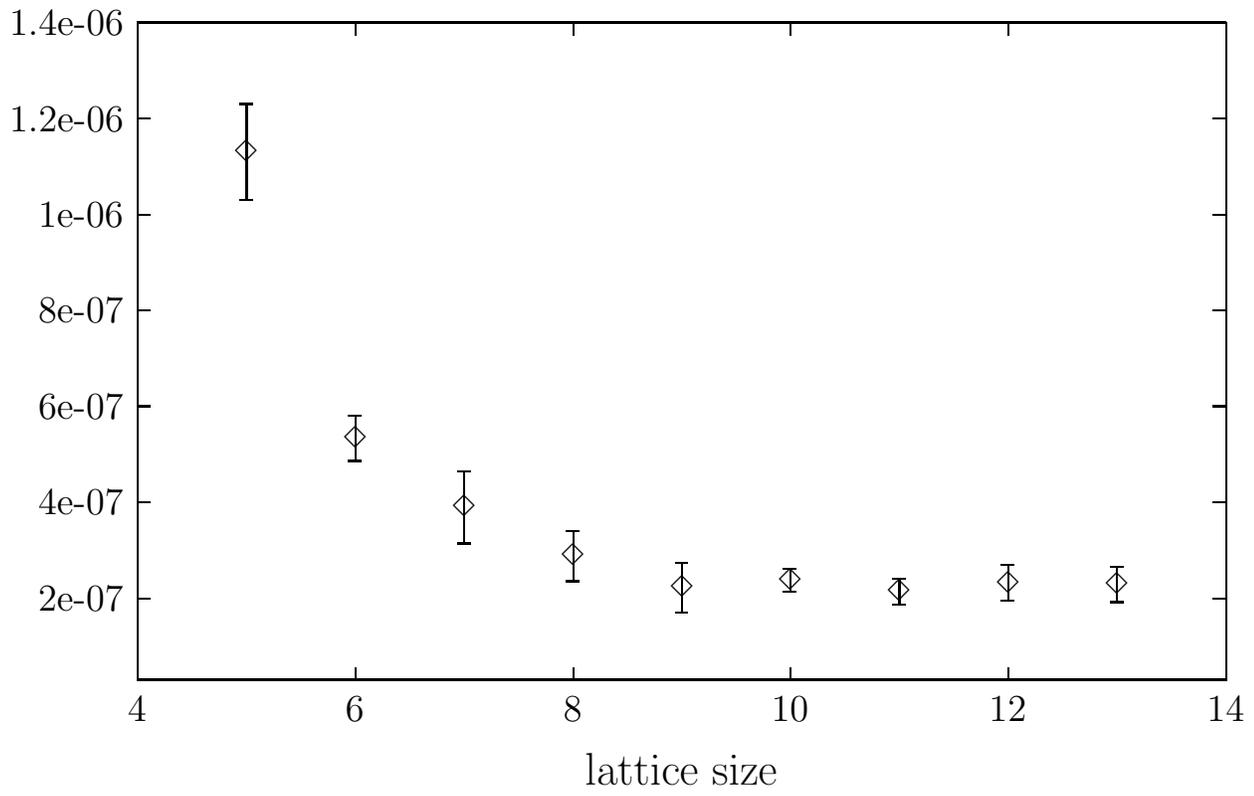
{\em Figure 2} & The probability of the vortex pair shown in Figure 1 as a
function of the lattice size at inverse temperature $\beta=2.0$. \\[2mm]

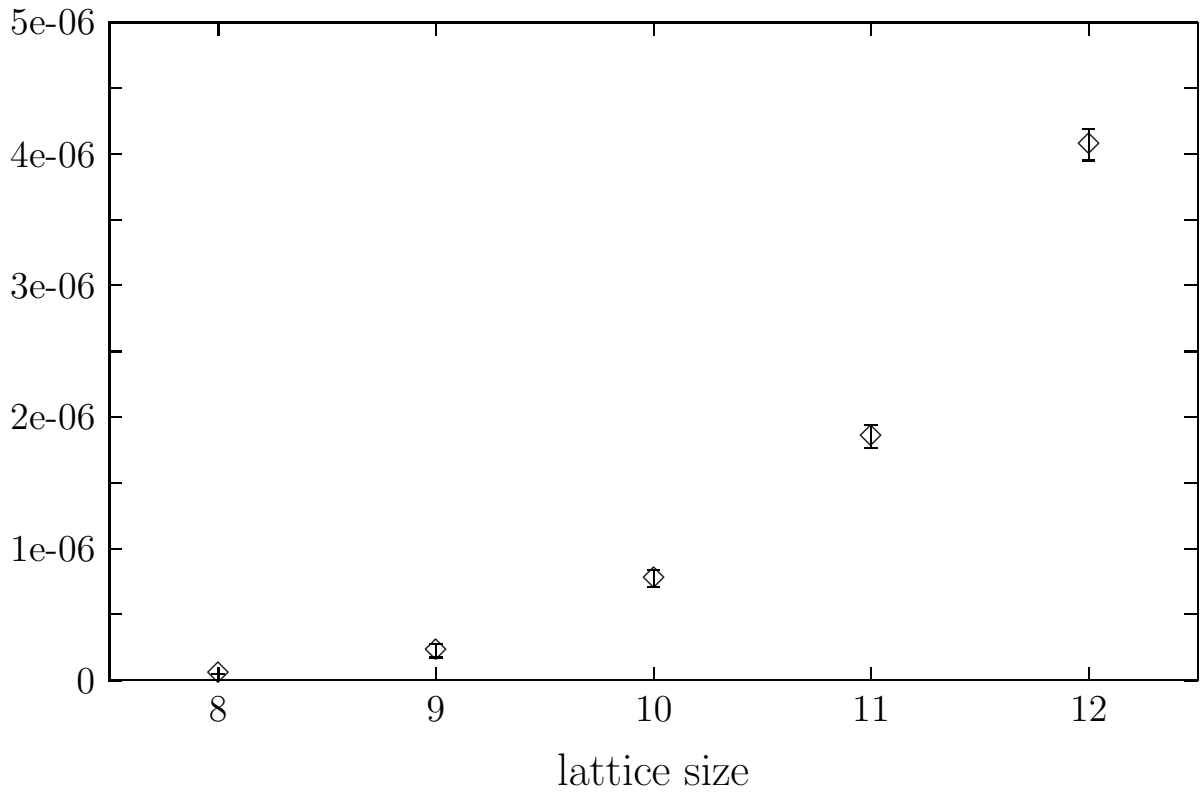
{\em Figure 3} & The probability of a vortex pair with $\eta$-string going
around the lattice at $\beta=2.0$. The length of the side of the lattice
parallel to the string is fixed to 9 and only the ``transverse'' size
(perpendicular to the string) is changed.

\end{tabular}
\pagebreak

\unitlength=1.00mm
\special{em:linewidth 0.4pt}
\linethickness{0.4pt}
\begin{picture}(110.00,159.00)
\put(10.00,30.00){\line(1,0){100.00}}
\put(110.00,30.00){\line(0,1){100.00}}
\put(110.00,130.00){\line(-1,0){100.00}}
\put(10.00,130.00){\line(0,-1){100.00}}
\linethickness{2.0pt}
\put(80.00,10.00){\line(1,0){10.00}}
\put(94.00,10.00){\makebox(0,0)[lc]{\Large $\eta_l=-1$}}
\put(0.00,159.00){\makebox(0,0)[lc]{{\LARGE \bf Figure 1}}}
\put(50.00,40.00){\line(1,0){10.00}}
\put(60.00,50.00){\line(-1,0){10.00}}
\put(50.00,60.00){\line(1,0){10.00}}
\put(50.00,80.00){\line(1,0){10.00}}
\put(60.00,90.00){\line(-1,0){10.00}}
\put(50.00,100.00){\line(1,0){10.00}}
\put(60.00,110.00){\line(-1,0){10.00}}
\put(50.00,120.00){\line(1,0){10.00}}
\put(60.00,130.00){\line(-1,0){10.00}}
\put(50.00,30.00){\line(1,0){10.00}}
\linethickness{0.4pt}
\put(10.00,70.00){\line(1,0){100.00}}
\put(90.00,74.00){\makebox(0,0)[cb]{{\Large $C$}}}
\put(55.00,75.00){\makebox(0,0)[cc]{{\Large $\otimes$}}}
\put(55.00,65.00){\makebox(0,0)[cc]{{\Large $\otimes$}}}
\put(25.00,10.00){\makebox(0,0)[lc]{{\Large $\otimes$ vortex}}}
\end{picture}
\pagebreak

% GNUPLOT: LaTeX picture
\setlength{\unitlength}{0.240900pt}
\ifx\plotpoint\undefined\newsavebox{\plotpoint}\fi
\sbox{\plotpoint}{\rule[-0.200pt]{0.400pt}{0.400pt}}%
\begin{picture}(1949,1169)(300,200)
\font\gnuplot=cmr10 at 10pt
\gnuplot
\sbox{\plotpoint}{\rule[-0.200pt]{0.400pt}{0.400pt}}%
\put(176.0,241.0){\rule[-0.200pt]{4.818pt}{0.400pt}}
\put(154,241){\makebox(0,0)[r]{\large 2e-07}}
\put(1865.0,241.0){\rule[-0.200pt]{4.818pt}{0.400pt}}
\put(176.0,392.0){\rule[-0.200pt]{4.818pt}{0.400pt}}
\put(154,392){\makebox(0,0)[r]{\large 4e-07}}
\put(1865.0,392.0){\rule[-0.200pt]{4.818pt}{0.400pt}}
\put(176.0,543.0){\rule[-0.200pt]{4.818pt}{0.400pt}}
\put(154,543){\makebox(0,0)[r]{\large 6e-07}}
\put(1865.0,543.0){\rule[-0.200pt]{4.818pt}{0.400pt}}
\put(176.0,694.0){\rule[-0.200pt]{4.818pt}{0.400pt}}
\put(154,694){\makebox(0,0)[r]{\large 8e-07}}
\put(1865.0,694.0){\rule[-0.200pt]{4.818pt}{0.400pt}}
\put(176.0,844.0){\rule[-0.200pt]{4.818pt}{0.400pt}}
\put(154,844){\makebox(0,0)[r]{\large 1e-06}}
\put(1865.0,844.0){\rule[-0.200pt]{4.818pt}{0.400pt}}
\put(176.0,995.0){\rule[-0.200pt]{4.818pt}{0.400pt}}
\put(154,995){\makebox(0,0)[r]{\large 1.2e-06}}
\put(1865.0,995.0){\rule[-0.200pt]{4.818pt}{0.400pt}}
\put(176.0,1146.0){\rule[-0.200pt]{4.818pt}{0.400pt}}
\put(154,1146){\makebox(0,0)[r]{\large 1.4e-06}}
\put(1865.0,1146.0){\rule[-0.200pt]{4.818pt}{0.400pt}}
\put(176.0,113.0){\rule[-0.200pt]{0.400pt}{4.818pt}}
\put(176,68){\makebox(0,0){\large 4}}
\put(176.0,1126.0){\rule[-0.200pt]{0.400pt}{4.818pt}}
\put(518.0,113.0){\rule[-0.200pt]{0.400pt}{4.818pt}}
\put(518,68){\makebox(0,0){\large 6}}
\put(518.0,1126.0){\rule[-0.200pt]{0.400pt}{4.818pt}}
\put(860.0,113.0){\rule[-0.200pt]{0.400pt}{4.818pt}}
\put(860,68){\makebox(0,0){\large 8}}
\put(860.0,1126.0){\rule[-0.200pt]{0.400pt}{4.818pt}}
\put(1201.0,113.0){\rule[-0.200pt]{0.400pt}{4.818pt}}
\put(1201,68){\makebox(0,0){\large 10}}
\put(1201.0,1126.0){\rule[-0.200pt]{0.400pt}{4.818pt}}
\put(1543.0,113.0){\rule[-0.200pt]{0.400pt}{4.818pt}}
\put(1543,68){\makebox(0,0){\large 12}}
\put(1543.0,1126.0){\rule[-0.200pt]{0.400pt}{4.818pt}}
\put(1885.0,113.0){\rule[-0.200pt]{0.400pt}{4.818pt}}
\put(1885,68){\makebox(0,0){\large 14}}
\put(1885.0,1126.0){\rule[-0.200pt]{0.400pt}{4.818pt}}
\put(176.0,113.0){\rule[-0.200pt]{411.698pt}{0.400pt}}
\put(1885.0,113.0){\rule[-0.200pt]{0.400pt}{248.850pt}}
\put(176.0,1146.0){\rule[-0.200pt]{411.698pt}{0.400pt}}
\put(1030,23){\makebox(0,-100){\Large lattice size}}
\put(176.0,113.0){\rule[-0.200pt]{0.400pt}{248.850pt}}
\put(347,942){\raisebox{-.8pt}{\makebox(0,0){$\Diamond$}}}
\put(518,492){\raisebox{-.8pt}{\makebox(0,0){$\Diamond$}}}
\put(689,384){\raisebox{-.8pt}{\makebox(0,0){$\Diamond$}}}
\put(860,308){\raisebox{-.8pt}{\makebox(0,0){$\Diamond$}}}
\put(1031,258){\raisebox{-.8pt}{\makebox(0,0){$\Diamond$}}}
\put(1201,269){\raisebox{-.8pt}{\makebox(0,0){$\Diamond$}}}
\put(1372,252){\raisebox{-.8pt}{\makebox(0,0){$\Diamond$}}}
\put(1543,265){\raisebox{-.8pt}{\makebox(0,0){$\Diamond$}}}
\put(1714,263){\raisebox{-.8pt}{\makebox(0,0){$\Diamond$}}}
\put(347.0,867.0){\rule[-0.200pt]{0.400pt}{36.376pt}}
\put(337.0,867.0){\rule[-0.200pt]{4.818pt}{0.400pt}}
\put(337.0,1018.0){\rule[-0.200pt]{4.818pt}{0.400pt}}
\put(518.0,457.0){\rule[-0.200pt]{0.400pt}{17.104pt}}
\put(508.0,457.0){\rule[-0.200pt]{4.818pt}{0.400pt}}
\put(508.0,528.0){\rule[-0.200pt]{4.818pt}{0.400pt}}
\put(689.0,327.0){\rule[-0.200pt]{0.400pt}{27.222pt}}
\put(679.0,327.0){\rule[-0.200pt]{4.818pt}{0.400pt}}
\put(679.0,440.0){\rule[-0.200pt]{4.818pt}{0.400pt}}
\put(860.0,268.0){\rule[-0.200pt]{0.400pt}{19.031pt}}
\put(850.0,268.0){\rule[-0.200pt]{4.818pt}{0.400pt}}
\put(850.0,347.0){\rule[-0.200pt]{4.818pt}{0.400pt}}
\put(1031.0,219.0){\rule[-0.200pt]{0.400pt}{18.790pt}}
\put(1021.0,219.0){\rule[-0.200pt]{4.818pt}{0.400pt}}
\put(1021.0,297.0){\rule[-0.200pt]{4.818pt}{0.400pt}}
\put(1201.0,251.0){\rule[-0.200pt]{0.400pt}{8.672pt}}
\put(1191.0,251.0){\rule[-0.200pt]{4.818pt}{0.400pt}}
\put(1191.0,287.0){\rule[-0.200pt]{4.818pt}{0.400pt}}
\put(1372.0,231.0){\rule[-0.200pt]{0.400pt}{9.877pt}}
\put(1362.0,231.0){\rule[-0.200pt]{4.818pt}{0.400pt}}
\put(1362.0,272.0){\rule[-0.200pt]{4.818pt}{0.400pt}}
\put(1543.0,237.0){\rule[-0.200pt]{0.400pt}{13.731pt}}
\put(1533.0,237.0){\rule[-0.200pt]{4.818pt}{0.400pt}}
\put(1533.0,294.0){\rule[-0.200pt]{4.818pt}{0.400pt}}
\put(1714.0,235.0){\rule[-0.200pt]{0.400pt}{13.490pt}}
\put(1704.0,235.0){\rule[-0.200pt]{4.818pt}{0.400pt}}
\put(1704.0,291.0){\rule[-0.200pt]{4.818pt}{0.400pt}}
\put(200.0,1600.0){\makebox(0,0)[lc]{\LARGE \bf Figure 2}}
\end{picture}

\pagebreak

% GNUPLOT: LaTeX picture
\setlength{\unitlength}{0.240900pt}
\ifx\plotpoint\undefined\newsavebox{\plotpoint}\fi
\sbox{\plotpoint}{\rule[-0.200pt]{0.400pt}{0.400pt}}%
\begin{picture}(1949,1169)(300,200)
\font\gnuplot=cmr10 at 10pt
\gnuplot
\sbox{\plotpoint}{\rule[-0.200pt]{0.400pt}{0.400pt}}%
\put(176.0,112.0){\rule[-0.200pt]{4.818pt}{0.400pt}}
\put(154,112){\makebox(0,0)[r]{\large 0}}
\put(1865.0,112.0){\rule[-0.200pt]{4.818pt}{0.400pt}}
\put(176.0,216.0){\rule[-0.200pt]{4.818pt}{0.400pt}}
\put(1865.0,216.0){\rule[-0.200pt]{4.818pt}{0.400pt}}
\put(176.0,319.0){\rule[-0.200pt]{4.818pt}{0.400pt}}
\put(154,319){\makebox(0,0)[r]{\large 1e-06}}
\put(1865.0,319.0){\rule[-0.200pt]{4.818pt}{0.400pt}}
\put(176.0,422.0){\rule[-0.200pt]{4.818pt}{0.400pt}}
\put(1865.0,422.0){\rule[-0.200pt]{4.818pt}{0.400pt}}
\put(176.0,526.0){\rule[-0.200pt]{4.818pt}{0.400pt}}
\put(154,526){\makebox(0,0)[r]{\large 2e-06}}
\put(1865.0,526.0){\rule[-0.200pt]{4.818pt}{0.400pt}}
\put(176.0,629.0){\rule[-0.200pt]{4.818pt}{0.400pt}}
\put(1865.0,629.0){\rule[-0.200pt]{4.818pt}{0.400pt}}
\put(176.0,733.0){\rule[-0.200pt]{4.818pt}{0.400pt}}
\put(154,733){\makebox(0,0)[r]{\large 3e-06}}
\put(1865.0,733.0){\rule[-0.200pt]{4.818pt}{0.400pt}}
\put(176.0,836.0){\rule[-0.200pt]{4.818pt}{0.400pt}}
\put(1865.0,836.0){\rule[-0.200pt]{4.818pt}{0.400pt}}
\put(176.0,939.0){\rule[-0.200pt]{4.818pt}{0.400pt}}
\put(154,939){\makebox(0,0)[r]{\large 4e-06}}
\put(1865.0,939.0){\rule[-0.200pt]{4.818pt}{0.400pt}}
\put(176.0,1043.0){\rule[-0.200pt]{4.818pt}{0.400pt}}
\put(1865.0,1043.0){\rule[-0.200pt]{4.818pt}{0.400pt}}
\put(176.0,1146.0){\rule[-0.200pt]{4.818pt}{0.400pt}}
\put(154,1146){\makebox(0,0)[r]{\large 5e-06}}
\put(1865.0,1146.0){\rule[-0.200pt]{4.818pt}{0.400pt}}
\put(347.0,113.0){\rule[-0.200pt]{0.400pt}{4.818pt}}
\put(347,68){\makebox(0,0){\large 8}}
\put(347.0,1126.0){\rule[-0.200pt]{0.400pt}{4.818pt}}
\put(689.0,113.0){\rule[-0.200pt]{0.400pt}{4.818pt}}
\put(689,68){\makebox(0,0){\large 9}}
\put(689.0,1126.0){\rule[-0.200pt]{0.400pt}{4.818pt}}
\put(1031.0,113.0){\rule[-0.200pt]{0.400pt}{4.818pt}}
\put(1031,68){\makebox(0,0){\large 10}}
\put(1031.0,1126.0){\rule[-0.200pt]{0.400pt}{4.818pt}}
\put(1372.0,113.0){\rule[-0.200pt]{0.400pt}{4.818pt}}
\put(1372,68){\makebox(0,0){\large 11}}
\put(1372.0,1126.0){\rule[-0.200pt]{0.400pt}{4.818pt}}
\put(1714.0,113.0){\rule[-0.200pt]{0.400pt}{4.818pt}}
\put(1714,68){\makebox(0,0){\large 12}}
\put(1714.0,1126.0){\rule[-0.200pt]{0.400pt}{4.818pt}}
\put(176.0,113.0){\rule[-0.200pt]{411.698pt}{0.400pt}}
\put(1885.0,113.0){\rule[-0.200pt]{0.400pt}{248.850pt}}
\put(176.0,1146.0){\rule[-0.200pt]{411.698pt}{0.400pt}}
\put(1030,23){\makebox(0,-100){\Large lattice size}}
\put(176.0,113.0){\rule[-0.200pt]{0.400pt}{248.850pt}}
\put(347,122){\raisebox{-.8pt}{\makebox(0,0){$\Diamond$}}}
\put(689,158){\raisebox{-.8pt}{\makebox(0,0){$\Diamond$}}}
\put(1031,272){\raisebox{-.8pt}{\makebox(0,0){$\Diamond$}}}
\put(1372,495){\raisebox{-.8pt}{\makebox(0,0){$\Diamond$}}}
\put(1714,954){\raisebox{-.8pt}{\makebox(0,0){$\Diamond$}}}
\put(347,122){\usebox{\plotpoint}}
\put(337.0,122.0){\rule[-0.200pt]{4.818pt}{0.400pt}}
\put(337.0,122.0){\rule[-0.200pt]{4.818pt}{0.400pt}}
\put(689.0,148.0){\rule[-0.200pt]{0.400pt}{5.059pt}}
\put(679.0,148.0){\rule[-0.200pt]{4.818pt}{0.400pt}}
\put(679.0,169.0){\rule[-0.200pt]{4.818pt}{0.400pt}}
\put(1031.0,259.0){\rule[-0.200pt]{0.400pt}{6.504pt}}
\put(1021.0,259.0){\rule[-0.200pt]{4.818pt}{0.400pt}}
\put(1021.0,286.0){\rule[-0.200pt]{4.818pt}{0.400pt}}
\put(1372.0,477.0){\rule[-0.200pt]{0.400pt}{8.672pt}}
\put(1362.0,477.0){\rule[-0.200pt]{4.818pt}{0.400pt}}
\put(1362.0,513.0){\rule[-0.200pt]{4.818pt}{0.400pt}}
\put(1714.0,929.0){\rule[-0.200pt]{0.400pt}{12.045pt}}
\put(1704.0,929.0){\rule[-0.200pt]{4.818pt}{0.400pt}}
\put(1704.0,979.0){\rule[-0.200pt]{4.818pt}{0.400pt}}
\put(200.0,1600.0){\makebox(0,0)[lc]{\LARGE \bf Figure 3}}
\end{picture}

\end{document}